\documentclass[aps,prc,twocolumn,twoside,showkeys,showpacs]{revtex4}
\usepackage{graphics,epsfig}

\newcommand{\figart}[5]{\begin{figure}[#1]
			\begin{center}
                        \includegraphics[scale=#5,draft=false]{#2.eps}
                        \caption{\small \it #3}
                        \label{#4}
                        \end{center}
                        \end{figure}}
			
\newcommand{\figartlarg}[5]{\begin{figure*}[#1]
			\begin{center}
                        \includegraphics[scale=#5,draft=false]{#2.eps}
                        \caption{\small \it #3}
                        \label{#4}
                        \end{center}
                        \end{figure*}}

\newcommand{\LPC}[0]{\it LPC Caen (IN2P3-CNRS/ISMRA et Universit\'e),
14050 Caen Cedex , France}

\newcommand{\potentiel}[0]{figure_1}
\newcommand{\staticvsn}[0]{figure_2}
\newcommand{\collisions}[0]{figure_3_lanl}
\newcommand{\wilczynski}[0]{figure_4}
\newcommand{\EvsV}[0]{figure_5}
\newcommand{\EvsN}[0]{figure_6}
\newcommand{\dalitzcnbd}[0]{figure_7_lanl}
\newcommand{\dalitzsmm}[0]{figure_8_lanl}

\begin{document}



\title{\Large \bf Selected aspects of the classical molecular dynamics of
$N$ bodies in strong interaction.}
\author{D.~Cussol}
\affiliation{\LPC}
\date{\today}

\begin{abstract}
Selected results of a classical simulation of $N$ bodies in strong interaction
are presented. The static properties of such classical systems are qualitatively
similar to the known properties of atomic nuclei. The simulations of collisions
show that all observed reaction mechanisms in nucleus-nucleus
collisions are present in this numerical simulation. The first studies of such
collisions are in qualitative agreement with experimental observations. This
simulation could shed a new light on energy deposit in heavy ion
collisions: the excitation energy of each cluster is found lower than 
the energy  of the less
bound body in the cluster. Finally, the similarities of fragmentation pattern 
with those
obtained with a statistical code indicate that an unambiguous link can be
established between the statistical and the dynamical descriptions of reaction
mechanisms.  
\end{abstract}

\maketitle



During the last decades, heavy ions collisions have been widely used and 
intensively
studied experimentally to determine the equation of state of nuclear matter. 
Many reaction
mechanisms have been identified and a lot of theoretical works has been induced.
Schematically, the incident energy domain has been divided in three parts: 
the low
energy range, up to 10$\sim$20 A.MeV, where fusion and deep inelastic processes
are dominant; the high energy range, above 100 A.MeV, where the nucleonic and 
sub-nucleonic
degrees of freedom start to play a significant role; and the intermediate energy
range, from 20 A.MeV to 100 A.MeV, where binary processes and multi-fragmentation
take place.
At each energy range, the reaction is schematically described as a two
steps process: an entrance channel during which excited nuclei are formed, 
and a decay stage where the excited fragments cool down by particle or fragment
emissions. 

Theoretically, the entrance channel of the reaction 
is mainly described by dynamical models. At low
energy, the attractive character of the interaction 
is dominant and the entrance channel is
described mainly by mean-field approaches \cite{TDHF}. 
At high energies, the repulsive part
of the interaction and its nucleon-nucleon character are dominant and cascade
models are mainly used to describe the first moments of the collision
\cite{cascadeCug}. At
intermediate energies, the attractive part and the repulsive part of the
interaction interfere. Transport models are widely used to describe the entrance
channel at this energy range \cite{LV,BNV,BUU,BLV,QMD,AMD,FMD}. 
They contain a mean-field part through the
one-body evolution of the system, and a nucleon-nucleon part through the
collision term.      
At all energy ranges, the decay stage is mainly described by statistical 
decay models (GEMINI \cite{GEMINI}, SIMON \cite{SIMON},
SMM \cite{SMM}, MMMC \cite{MMMC}, QSM \cite{QSM}, EES \cite{EES}, etc.) 
which assume 
that the nuclei formed during the first dynamical step are equilibrated.

With this scheme, it is very hard to have a global and consistent 
description of
nucleus-nucleus collisions. Although the nucleon-nucleon interaction is the
same, whatever the incident energy and whatever the reaction time, different
models are used depending on the energy range and the reaction time. The links
between these different models are not obvious. For example how the liquid-drop
parametrisation used in the statistical decay code can be deduced from
the parameters of the interaction used in a transport code? 
What is the link between the
incompressibility modulus of nuclear matter $K_\infty$ and the nucleon-nucleon
cross section $\sigma_{nn}$? Is the statistical description still justified when
the thermalisation time is close to or longer than the life time of 
excited nuclei?

The present status results mainly from the impossibility to solve 
the full nuclear many-body problem. To solve it, approximations are made,
depending on the energy range and on the reaction time. Even after
simplifications, the resulting equations do not 
have analytical solutions and they are
often solved numerically. It seems hard nowadays to connect these different
models to have a global description of the processes involved in nucleus-nucleus
collisions.    

Nevertheless, some attempts were made to describe globally nucleus-nucleus
reactions at intermediate energies by using (semi-)classical molecular
dynamics codes 
\cite{Pandhar87,Dorso88,Latora94,Bondorf97,Sator2001,PlagnolBormio,Dorso99}. 
The main advantage of these codes is that the classical
many-body problem can be solved numerically with a high accuracy without any
assumption. The main drawback is that the quantum character of the system is
ignored. 
A lot of work has already been done with the help of such codes
to study the mechanism of fragment formation in multi-fragmentation. Could these
codes do better by describing the whole reaction process, from the very
beginning of the reaction up to the decay stage? Are they able to make a link
between the different energy ranges? Which processes are accessible with
the simplest hypothesis?

This paper will show that
classical molecular dynamics simulations are interesting tools to establish 
the links
between the different approaches used in nucleus-nucleus collision. In the first
section, the Classical N-Body Dynamics code will be briefly described. The
static properties of stable N-body systems will be studied in the second
section. Some examples of reaction mechanisms and
some analyses of cluster-cluster collisions are given in the 
third section. Finally, conclusions will be drawn. 

\figart{ht}{\potentiel}
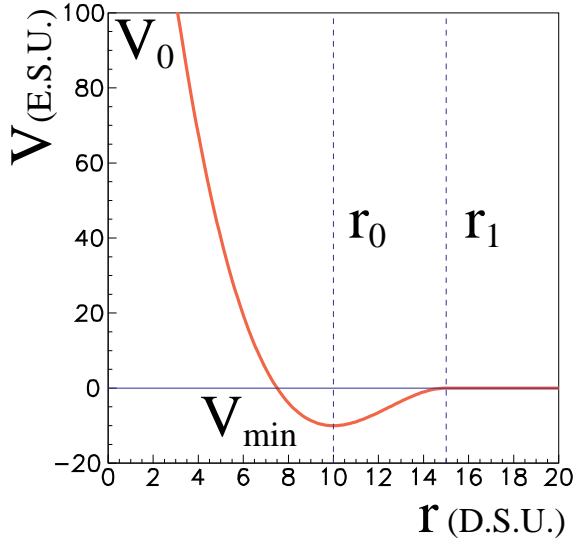
{Shape of the interaction used in the CNBD
simulation}{f:potentiel}{0.4}

\section{Description of the code.}
Let us start by describing the Classical N-Body Dynamics code (labeled CNBD)
used in this article.
The basic ingredients of such a code are very simple. The dynamical evolution of 
each body of
the system is driven by the classical Newtonian equations of motion. The
two-body potential used in the present work is a third degree polynomial whose
derivatives are null at the range $r_1$ and at the distance of maximum depth
$r_0$. The depth value is $V_{min}$ and the value at $r=0$ is finite and equal 
to $V_0$. The shape of the two-body potential is shown on figure
\ref{f:potentiel}. This potential has the basic properties of the Lennard-Jones
potential used in other works \cite{PlagnolBormio,Dorso99}: 
a finite range attractive part and a repulsive
short range part. To follow the dynamical evolution of the system 
an adaptative stepsize fourth-order Runge-Kutta algorithm is used 
\cite{NumRec}. The main difference with other works is that the time step
$\Delta t$ can vary: if the potential varies strongly, $\Delta t$ is small and
when the potential varies gently, $\Delta t$ becomes larger. 
This allows a very high
accuracy with shorter CPU time than for fixed time step algorithms.
It requires an additional simulation parameter $\epsilon$ which is adjusted 
to ensure the
verification of conservation laws (energy, momentum, angular momentum) with a
reasonable simulation time. 
For an $\epsilon$ value of $10^{-5}$, typical CPU times for a collision of two
clusters with 50 particles each with an ending time equal to 200 time 
simulations units 
on a Compaq DS20 computer under the UNIX True64 operating
system ranges from $\approx 30$ to $\approx 400$ seconds, depending mainly on 
the impact parameter. The energy gap between the beginning and the
the ending simulation time is lower than 0.001\%.
This simulation has five free parameters: four
linked to the physics (the interaction) and one linked to the numerical
algorithm ($\epsilon$).


Since one wants to study the simplest case, neither long range repulsive
interaction nor quantum corrections like a Pauli potential have been introduced
\cite{Dorso87}.
Additionally, no statistical decay code is applied on the excited fragments 
formed during the collision. The final products have to be regarded as
``primary'' products which will decay afterwards. 

In order to avoid any confusion with nuclear physics, the units used here are
called Simulation Units and noted $S.U.$. The distance will then be in Distance
Simulation Units ($D.S.U.$), the energies in Energy Simulation Units ($E.S.U.$),
the velocities in Velocity Simulation Units ($V.S.U.$) and the reaction time in
Time Simulation Unit ($T.S.U.$). We will be only interested in the relative
evolutions of the observables and in their link to the properties of the stable
systems. The main goal of the present work is not to reproduce the 
experimental data of nucleus-nucleus collisions, but rather to see to 
which extent
this simple simulation is qualitatively similar, or not, to experimental
data.     

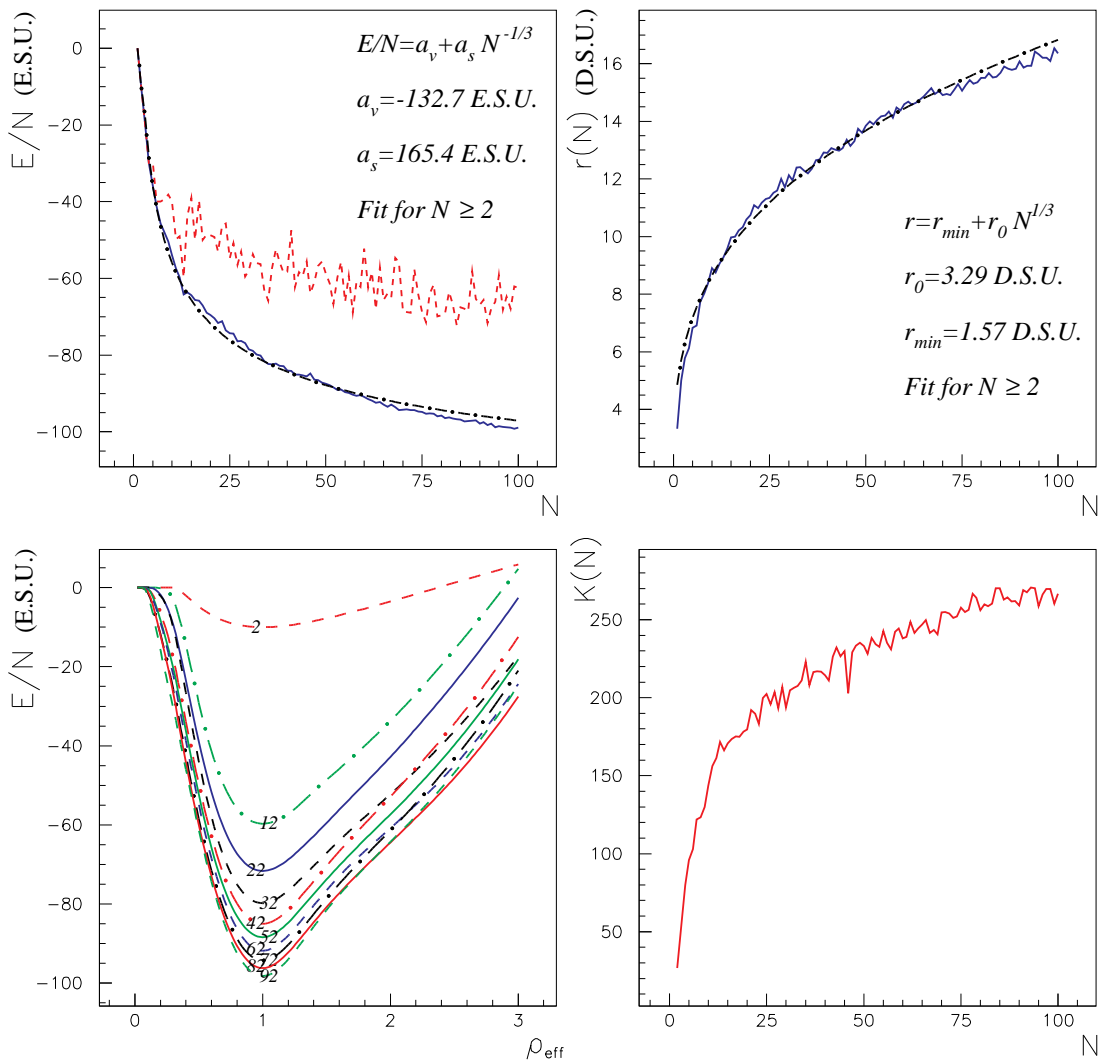
\figartlarg{t}{\staticvsn}{Energy per body (upper left panel) and root mean
square radius (upper
right panel) of N-body clusters as functions of $N$. For these two panels, the
full line correspond to the obtained values and the dashed and dotted line to a
fit. On the left panel, the dashed line corresponds to the energy of the less
bound body in the cluster. Lower row: zero
temperature ``equation of state'' for different $N$ values (left) and
``incompressibility modulus'' as a function of $N$ (right).}
{f:static_prop}{0.46}

\section{Static properties of ``ground states''.}
Once the basic ingredients are defined, one can build stable systems. Since the
two-body potential only depends on the distance between the two bodies, such
systems are small crystals. The ``ground states'' of such systems are defined 
as the configuration in position space which minimizes their total energy. 
This is obtained by using a Metropolis simulated annealing method
\cite{NumRec}. The locations of particles obtained this way are very close to
those obtained by using a basin-hopping algorithm for Lennard-Jones clusters 
\cite{WALES97}.

On top-left of figure \ref{f:static_prop} is displayed the energy per body
$E_{Bind}/N$
of these ``ground states'' as a function of the number $N$ 
of bodies in the cluster (solid line).
The dashed and dotted line corresponds to a fit using a liquid-drop formula. 
The dashed line corresponds to the energy of the less bound body 
$E_{LessBound}$ in the cluster. It will be seen that 
this energy seems to play a particular role
in cluster-cluster collisions. The overall dependence of $E_{Bind}/N$ with $N$ 
is very similar to what is seen in nuclear physics. The main difference is seen
for high values of $N$ because of the absence of a Coulomb-like interaction:
$E_{Bind}/N$ continues to decrease with $N$ whereas it increases for nuclei.
These values are close to those found by using the basin-hopping algorithm
\cite{WALES97}.


The dependence with $N$ of the root mean radius radius 
$r(N) = \sqrt{<r^2>}$ of N-body clusters is shown on the
top-right panel of figure \ref{f:static_prop} (solid line). 
The dashed and dotted line
corresponds to a fit using a $N^{\frac{1}{3}}$ term. Here again, this is similar
to what is known for nuclei. The main difference is the $r_0$ term which is due
to the large size of the ``repulsive core'' compared to the range of the
attractive  part of the interaction.

Since a size can be defined, an effective density can be calculated. One can
stretch and squeeze the N-body cluster and build an effective ``zero
temperature equation of state''. This is displayed on the low-left panel on
figure \ref{f:static_prop} for various system sizes. The density $\rho_{eff}$ on
this abscissa is the density relative to the density for the ``ground state''.
The curvature of this ``equation of state'' can be computed and defined as the
``compressibility modulus'' $K(N)$ of the system. Its variations with $N$ are 
displayed
on the low-right panel of figure \ref{f:static_prop}. One can see on these two
plots that the ``equation of state'' and $K(N)$ are strongly dependent on $N$
for values of $N$ below 20$\sim$30 and then less dependent on $N$ above. One
could be able to define an ``equation of state of infinite matter'' by computing
the limit of these evolutions for huge values of $N$. Here again, the behavior
of these classical N-body systems is very similar to that of nuclei.

This first study shows that the classical 
N-body clusters have strong similarities 
with atomic nuclei.
This suggests that all the parametrisation used to describe nuclei 
(liquid
drop parametrisation, radii, equation of state, etc...) could be directly 
deduced from the parameters of the two-body interaction. 
Since this interaction
remains unchanged, the differences seen for different clusters has to be
attributed only to $N$ and to the geometrical configuration of the bodies 
in the clusters.

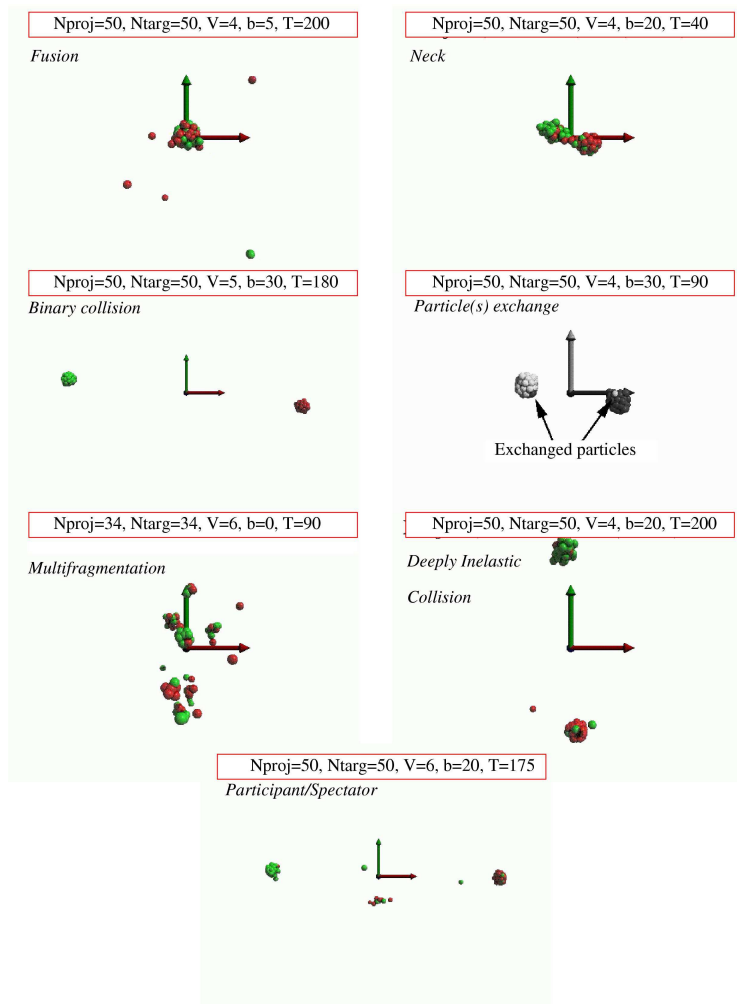
\figartlarg{ht}{\collisions}{Pictures of cluster-cluster collisions for different 
reaction mechanisms. On each panel, $Nproj$ is the projectile size, $Ntarg$ the
target size, $V$ the relative velocity in $V.S.U.$ 
between the target and the projectile , $b$ the impact parameter
in $D.S.U.$ and $t$ the collision time in $T.S.U.$.}{f:coll}{0.50}

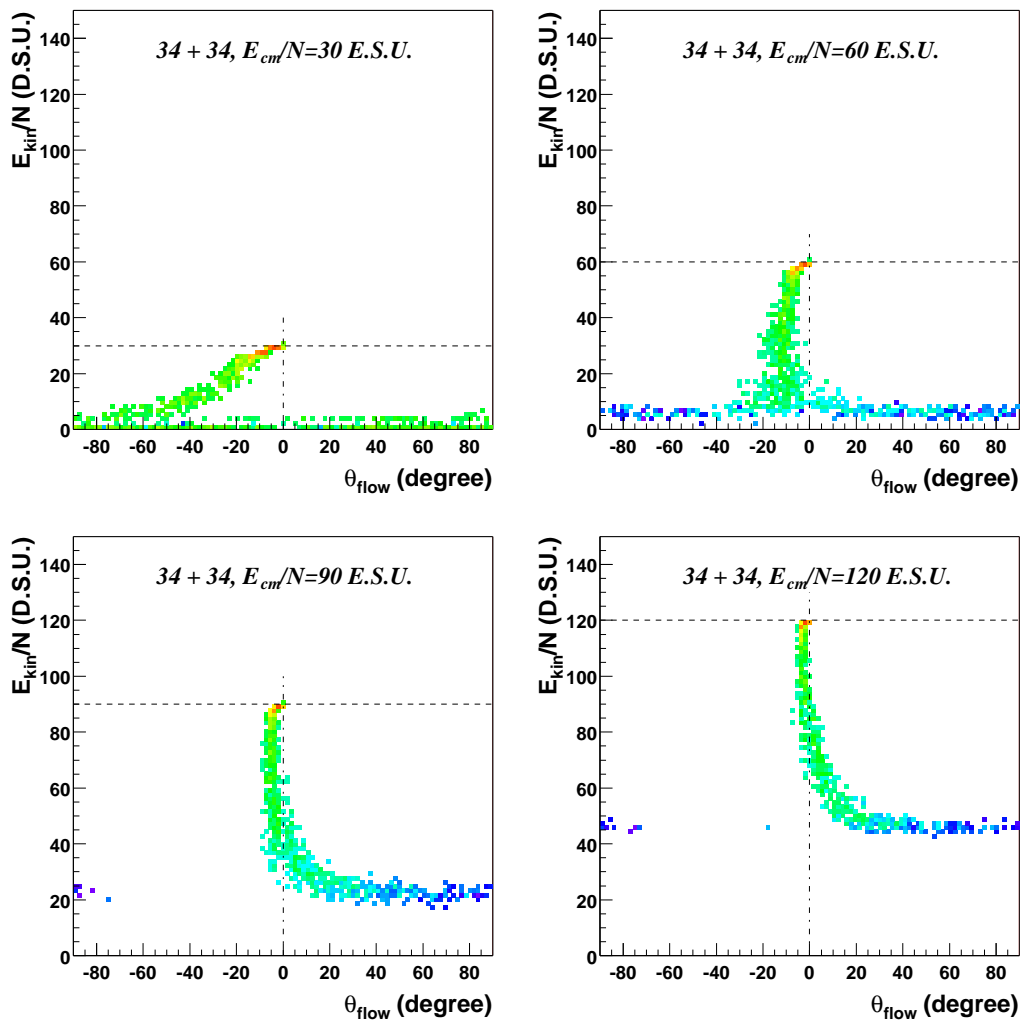
\figartlarg{ht}{\wilczynski}{Wilczinky plots for $N=34$ + $N=34$ collisions at
different available energies in the center of mass (see text).}
{f:Wilczynski}{0.7}

\section{Cluster-cluster collisions.}
After studying the static properties of stable N-body systems, collisions
between clusters can be investigated 
to see what kind of reaction can be obtained. 
Roughly 20,000 collisions have been generated for different system sizes,
different entrance channel asymmetries and different impact parameters. For each
collision, the orientation of the inertia axis of clusters are randomly chosen.
This avoid to run twice the same collision if the impact parameter, the target
and the projectile size and the incident energy are the same for two different
collisions. The simulations have been performed for the following 
systems: 13~+~13, 34~+~34, 
50~+~50, 100~+~100 and 18~+~50. The incident energies have been chosen in a way 
that the
available energy in the center of mass frame ranges from an energy far
below the 
binding energy of the fused system ($\approx 30 \> E.S.U$) to an 
energy 
well above the binding
energy of the fused system ($\approx 120 \> E.S.U.$).
As
it will be shown, these collisions can be studied in the same way as the
experimental data of nucleus-nucleus collisions are. These collisions can be
seen as ``numerical experiments''. Only few examples of such
analyses will be shown. More detailed analyses will be done in forthcoming
articles. Firstly, a small list of reaction mechanisms obtained 
will be done. Secondly, the excitation energy stored in excited clusters 
will be
studied and a link to the static properties will be done. Finally, the question
of the statistical description of this dynamical model will be briefly 
addressed.

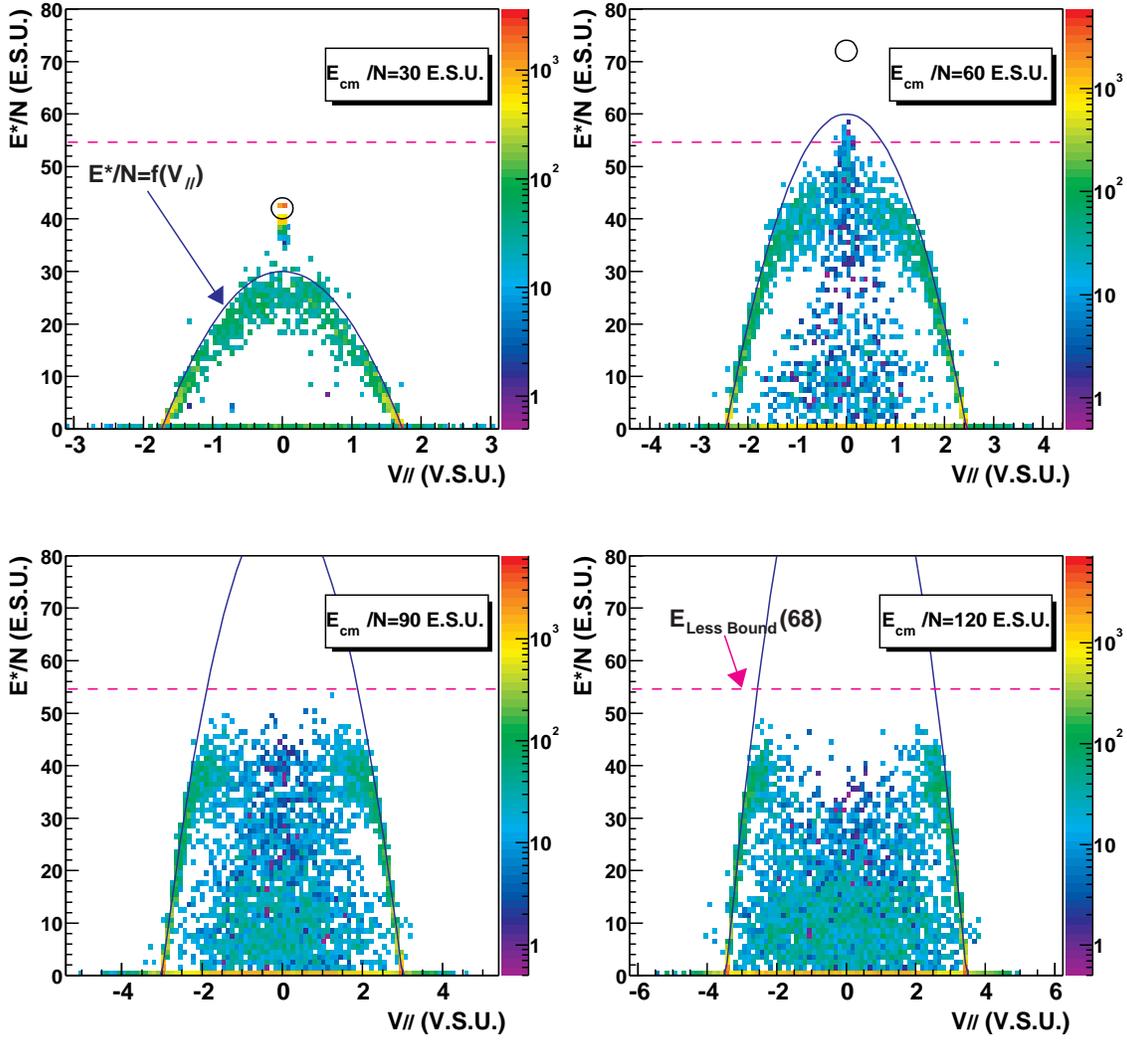
\figartlarg{ht}{\EvsV}{Excitation energy of clusters versus their parallel 
velocity for $N=34$ + $N=34$ collisions at
different available energies in the center of mass. On each panel, 
the full line corresponds to
the expected evolution for a pure binary process, 
the dashed horizontal line
corresponds to the energy of the less bound body for $N=68$ and 
the circle corresponds
to the expected values for the fused system.}
{f:EvsV}{0.75}

\figartlarg{ht}{\EvsN}{Excitation energy of clusters versus $N$ 
for $N=34$ + $N=34$ collisions at
different available energies in the center of mass. On each panel, 
the full line corresponds to
the energy of the less bound body, 
the dashed line
corresponds to the binding energy per body and 
the circle corresponds
to the expected values for the fused system.}
{f:EvsN}{0.75}

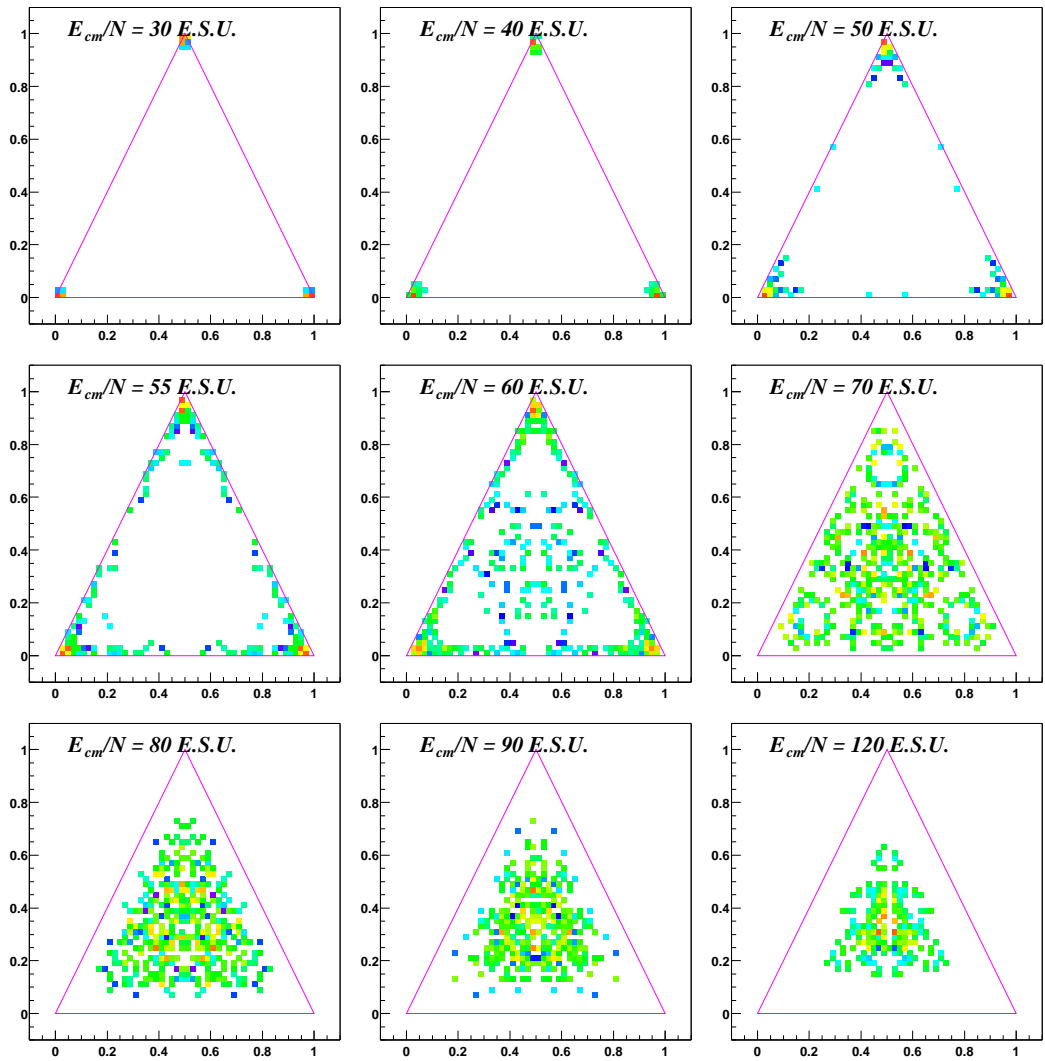
\figartlarg{t}{\dalitzcnbd}{Dalitz plots for CNBD simulations for central 
($0 \leq b/b_{max} < 0.1$) $N=34$ + $N=34$ collisions at
different available energies in the center of mass.}
{f:dalitz_CNBD}{0.70}

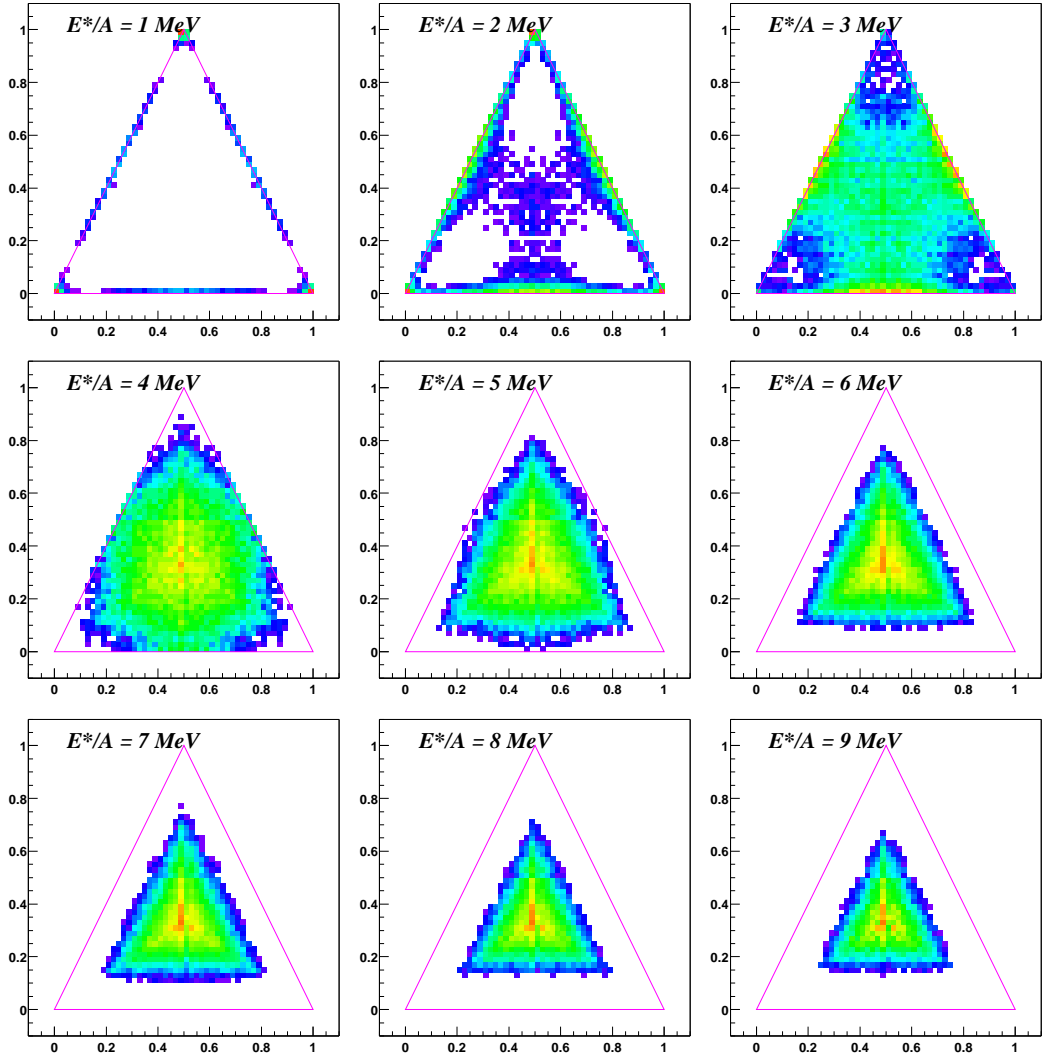
\figartlarg{t}{\dalitzsmm}{Dalitz plots obtained with SMM calculations for
different excitation energies. The source has a charge $Z=86$, a mass $A=205$
and the freeze-out density is $\rho_{f.o.}=\rho_0/3$.}
{f:dalitz_smm}{0.70}

\subsection{Reaction mechanisms.}
On figure \ref{f:coll} are displayed the pictures of cluster-cluster collisions 
obtained at a fixed time. 
The ending time of these collisions is $t=200 \> T.S.U.$.
On each picture, the bodies which were originally belonging to the projectile
are in dark grey and those which were belonging to the target are in light grey.
The most striking observation is that all kinds of reaction mechanisms observed
in nucleus-nucleus collisions seem to be present in this very simple
simulation. One can see low energy processes like the fusion/evaporation process
(first row left panel), pure binary collision (second row left panel), 
stripping/pick-up
mechanism (second row right panel) and deep inelastic collisions 
(third row right panel).
Intermediate energy processes, like neck formation and break-up (first row
right panel) and multi-fragmentation (third row left panel), are seen. 
Finally, high energy processes like the participant/spectator process 
(lower-most panel) are also seen.

This similarity can also be seen on the so-called Wilczynski plots
\cite{Wilczynski} shown on figure
\ref{f:Wilczynski}. These plots display the correlation between the flow angle
$\theta_{flow}$ and the total kinetic energy of the clusters $E_{kin} =
\sum{E_k^i}$ where $E_k^i$ is the kinetic energy of cluster $i$ in the center
of mass frame. Two bodies are assumed to belong to the
same cluster if they are in interaction, i.e. if their relative distance is
below the range $r_1$ of the interaction (see figure \ref{f:potentiel}). 
This algorithm of cluster recognition is the
simplest one and is called Minimum Spanning Tree algorithm in other works
\cite{Dorso99,Dorso93}. 
The most
dissipative collisions 
correspond to the smallest $E_{kin}$ values: the
available energy is converted in internal energy of clusters. These
plots were built for a $N=34$ projectile colliding a $N=34$ target at four
available energies $E_{cm}/N$ in the center of mass: 
$E_{cm}/N=60 \> E.S.U.$, $E_{cm}/N=90 \> E.S.U.$, $E_{cm}/N=120 \> E.S.U.$ and 
$E_{cm}/N=30 \> E.S.U.$ which 
corresponds respectively to 
the energy of the less bound body for the fused $N=68$ cluster, 
to the binding energy per body for $N=68$, 
to the energy of the most bound body for $N=68$ and to 
an energy below the energy of the less bound body for $N=68$. For each energy,
1,000 collisions have been computed assuming a flat impact parameter
distribution ranging from $b=0 \> D.S.U.$ to the sum of the two cluster radii 
plus
the range of the interaction ($b_{max}=r(N_{proj})+r(N_{targ})+r_1$, 
where $r_1$ is
the range of the two-body interaction). In the
analyses, each collision is weighted assuming a triangular impact parameter
distribution between 0 and $b_{max}$ ($weight \propto b$).

At 
$E_{cm}/N=30 \> E.S.U.$ (upper left panel), the flow angle is always negative.
This means that the projectile and the target like clusters are deflected to
the opposite direction relative to their original one. At this energy, the
attractive part of the interaction is dominant. For the lowest $E_{kin}$ values,
all possible values of $\theta_{flow}$ are covered: this corresponds to the
fusion/evaporation process. At $E_{cm}/N=60 \> E.S.U.$ (upper right panel), 
the picture is 
changed and the range of $\theta_{flow}$ values is smaller than for the
previous energy: the repulsive part starts to 
act. The fusion/evaporation area
(small $E_{kin}$ values, all $\theta_{flow}$ values) is still present. For the
two highest energies (lower panels), the picture is roughly the same. For the
less dissipative collisions, the attractive part is still dominant (negative 
$\theta_{flow}$ values). But when the dissipation increases, the repulsive part
becomes dominant and the projectile and the target bounce on each other
(positive $\theta_{flow}$ values). For the most dissipative collisions, only
positive $\theta_{flow}$ values are seen which indicates the disappearance of
the fusion/evaporation process. These evolutions are qualitatively similar to
what is seen in nucleus-nucleus collisions 
(see for example \cite{Metivier2000}).

These two studies indicate that the description of the reaction mechanisms in 
terms of mean-field at low energies and in terms of 
nucleon-nucleon collision at
high energies can be deduced from the properties of the nucleon-nucleon
interaction only. The similarities of this classical N-body simulation with
nucleus-nucleus collisions suggest that, 
apart for the quantum mechanics effects,
an effective description of nucleus-nucleus collisions could be obtained with a
reduced number of parameters. Providing the interaction has a finite range
attractive part and a short range repulsive part, the overall behavior of the
N-body systems seems to be independent of the values of the parameters of this
interaction. For a more quantitative agreement with nucleus-nucleus
collisions, additional physical
ingredients (quantum mechanics, Coulomb interaction) have to be included.

\subsection{Energy deposit in clusters.}
Let us now focus our study of cluster-cluster collision on selected topics. Of
particular interest is the excitation energy stored in clusters. 
How this energy can be linked to the available energy and how it is linked to 
the properties of the ``ground state'' characteristics of these clusters?

One can plot for example the evolution of the excitation energy $E^*/N$  
of the cluster with its parallel velocity
$V_{//}$ for $N=34$ + $N=34$ collisions 
(figure \ref{f:EvsV}) for the whole impact parameter range. 
The excitation energy of each cluster is
simply the difference between the total energy (potential plus kinetic) and the
``ground state'' energy of the cluster. It reads:

\begin{equation}
E^*=\sum_{i=1}^{N}{E_{kin}^i}+\sum_{i,j>i}{V(r_{ij})}-E_{Bind}(N)
\end{equation}

\noindent where $E_{kin}^i$ is the kinetic energy of the body $i$ in the cluster
center of mass frame, $r_{ij}$ the relative distance between the bodies $i$ and
$j$, $V(r_{ij})$ the potential energy and $E_{Bind}(N)$ the binding energy of
the ``ground-state'' of the cluster with $N$ bodies.
This
excitation energy is determined at the end of the calculation corresponding to 
$t=200 \> T.S.U.$. This energy is very close to the one obtained at the 
separation
time of the clusters (the smallest time at which clusters can be identified), 
since in this time range the evaporation is very weak and
the clusters have no time to cool down significantly \cite{Dorso99}. 
On each panel of figure
\ref{f:EvsV}, the full line corresponds to the expected correlation between
$E^*/N$ and $V_{//}$ for a pure binary scenario (the excitation energy is only
due to the velocity damping of each partner), the horizontal dashed line to 
the energy
of the less bound body for the fused system $N=68$ and the small circle is
centered around the expected values of velocity and excitation energy for the
fused system.

At $E_{cm}/N=30 \> E.S.U.$, the points are slightly below the full line. This
means that the excitation energy is strongly linked to the velocity damping.
The small shift is due to mass transfers between the projectile and
the target, and to promptly emitted clusters. 
The area corresponding to the fused
system is well populated showing that a complete fusion process occurs. At 
$E_{cm}/N=60 \> E.S.U.$, the distribution of points is roughly compatible 
with the pure binary process hypothesis, but the complete fusion process area is
empty. The horizontal line seems to be an upper limit to the excitation energy
which can be stored in clusters. Clusters with rather small excitation energies
are found at velocities around the center of mass velocity. For the two highest
energies, this trend is enhanced: a set of clusters are around the full line, 
and when the full line is above the horizontal line, 
one finds clusters at small excitation
energies around the center of mass velocity. The energy of the less bound 
body
seems to be a limit to the excitation energy which can be stored in these
clusters.

This can be more clearly seen when the excitation energy $E^*/N$ is plotted 
as a function of $N$, 
as in figure \ref{f:EvsN}. On each panel, the full line corresponds to
the energy of the less bound body $E_{LessBound}$ in the cluster and the dashed
line to the binding energy per body $E_{bind}/N$. As in figure \ref{f:EvsV}, the
small circle corresponds to the expected values for the fused system. At 
$E_{cm}/N=30 \> E.S.U.$, the area corresponding to complete fusion is filled
and all the available energy can be stored as excitation energy. But for higher
energies, one can clearly see that for each fragment size, $E^*/N$ never
overcomes $E_{LessBound}$. At $E_{cm}/N=60 \> E.S.U.$ clusters with sizes
higher than the projectile size and the target size can be seen. This area
corresponds to an incomplete fusion process. For the two highest energies
($E_{cm}/N=90 \> E.S.U.$ and $E_{cm}/N=120 \> E.S.U.$), the
plots are almost identical: there is no more fusion and the clusters are smaller
than the target and than the projectile. 
One can notice that $E^*/N$ never reaches the
binding energy $E_{bind}/N$ except for small clusters where $E_{bind}/N$ and
$E_{LessBound}$ are equal. 

This limitation of excitation energy can be understood quite easily. 
The less bound body remains bound to the cluster only if its total
energy is negative, i.e. its kinetic energy due to the excitation 
is below its potential one. If one
assumes that the excitation energy is roughly equally shared over all bodies in
the cluster, when the kinetic 
energy balances the potential energy of the less bound body, 
this body is no more bound to the cluster and can escape. To be
observed for a long time, the excited cluster must have an excitation energy 
per body below the energy of the less bound body.

The mechanism of energy deposit in classical N-body clusters seems to be the
following one: the excitation seems to be mainly driven by the velocity damping
of the two partners and to a lesser extent by exchanges of 
bodies between them.
Once the energy of the less bound body is reached, unbound bodies and/or
clusters escape quickly and keep an excitation energy per body below the energy
of the less bound one. As a consequence, the highest energy deposit per body 
can only be
obtained at energies close to $E_{LessBound}$. For higher available energies,
the system fragments quickly, leaving rather ``cold'' clusters around the center
of mass velocity. This could be an
explanation to the quite low excitation energies of clusters found in central
collisions for the Xe + Sn system at 50 A.MeV \cite{Marie98}. This subject will
be more completely covered in a forthcoming article. 

\subsection{Statistical description of collisions.}
Let us end with fragment size analyses of cluster-cluster collisions. In
nucleus-nucleus collisions studies, such
analyses are very often used to fix the parameters of statistical models and to
verify the compatibility of statistical decay models with experimental data 
(see for example \cite{Marie97,DAgostino99a,Bocage}). 
The so-called Dalitz plot for central ($0 \leq b/b_{max} <
0.1$) $N=34$ + $N=34$
collisions are shown on figure \ref{f:dalitz_CNBD}. Each panel corresponds to a
fixed available energy in the center of mass ranging from 
$E_{cm}/N=30$ to  $120 \> E.S.U.$. For this analysis,
only the three heaviest clusters are taken into account. Each event is
associated with a point in this plot. The distance of one point with respect to
each edge of the triangle is proportional to the size of each of the three 
clusters. The corners 
correspond to events with a large size cluster and two
small ones (fusion/evaporation process), an event in the middle of an edge
corresponds to an event with two equal size clusters and a small one (fission
process or binary collision) and the center of the triangle corresponds to three
equal size clusters (multi-fragmentation process). One can see that when the
available energy increases, the reaction mechanisms goes continuously from
fusion/evaporation to binary collisions and finally to multi-fragmentation and/or
vaporization.

This picture is qualitatively very similar to the one obtained in 
SMM calculations \cite{SMM}, describing the decay of a single 
source, as shown on figure \ref{f:dalitz_smm}. In this case, 
each panel corresponds to an excitation energy. As in figure
\ref{f:dalitz_CNBD}, the lowest energies correspond to fusion/evaporation like
processes, and goes with a continuous transition towards fission and 
multi-fragmentation when the
excitation energy increases.

The strong qualitative similarity between these two pictures suggests that it
could be possible to build a statistical decay code which would give the same
results as the dynamical simulation. Of particular interest would be the study
of the relations between the parameters of the statistical decay code 
(size of the
source, excitation energy, deformation, radial flow, freeze-out volume, 
external constraints) with
the parameters of the dynamical one (projectile and target sizes, impact
parameter, parameters of the two-body interaction). This could also allow to
check under which conditions the statistical decay code can be applied. This
study could finally 
allow us to have a consistent description of cluster-cluster
collisions, where the parameters of the statistical decay 
model are deduced
from the parameters of the dynamical simulation. In this case, the
cluster-cluster collisions
could be described completely with a reduced number of parameters.

\section{Conclusions}
As it has been seen in this brief overview of the characteristics and the
reaction mechanisms of classical $N$ bodies in strong interaction, there are
strong qualitative similarities between these systems and the atomic nuclei. 
The static properties and the reaction mechanisms observed for the atomic 
nuclei and for these classical clusters are found very close to each other.
This
could mean that the experimental observations made for nucleus-nucleus 
collisions are mainly
governed by the N-body character of the system and by the overall shape of the
two-body interaction (finite range attractive part and short range repulsive
part). More physical ingredients (Coulomb interaction, quantum
mechanics) are of course necessary if one wants to have a
quantitative agreement with experimental data. 
But it is surprising to have such a qualitative 
agreement while essential physical ingredients are 
missing in the simulation. 

Since all kind of reaction mechanisms are observed in these classical
simulations, from the low energy fusion/evaporation processes to the high energy
participant/spectator processes, they may also allow to connect in a 
consistent way
the mean-field and the nucleon-nucleon approaches. One could for example study
the relation between the incompressibility modulus and the size of the repulsive
part of the interaction ($K_\infty$ and $\sigma_{nn}$ in the nuclear case). It
is well known in transport calculations that one can ``stiffen'' the effective
equation of state by just increasing the value of the nucleon-nucleon cross
section $\sigma_{nn}$ \cite{DLM92}.

Finally, such simulations may reconcile two approaches which were up to now
opposed in nuclear physics: the dynamical description and the statistical 
description of fragment emissions.  
The dynamical models are unique tools to establish the
link between the parameters of the interaction and those of the statistical 
models.
Additionally, one can determine under which conditions the statistical 
models can
be used, and what are the true meanings of the parameters used in statistical
models (temperature, chemical potentials, freeze-out volumes, etc.). This study
may allow to have a complete and consistent description of colliding N-body 
systems with a reduced number of free parameters. This could be not only useful
for nuclear physics, but also for other fields of physics like for example the
cluster physics.     

\section*{Acknowledgments}
I would like to thank my colleagues of the INDRA collaboration and from the
Laboratoire de Physique Corpusculaire de Caen for a careful reading of this
article. I especially thanks Denis Lacroix for suggestions and fruitful
discussions. I would like also to thank O.Lopez and A.Botvina for 
the SMM calculations.



\end{document}